\documentclass[twocolumn,aps,prl,superscriptaddress,showpacs]{revtex4}
\usepackage{graphicx}
\usepackage{amsmath}
\usepackage{amsfonts}
\usepackage{amssymb}
\begin{document}

\author{Yaroslav Tserkovnyak}
\affiliation{Lyman Laboratory of Physics, Harvard University,
 Cambridge, Massachusetts 02138, USA}

\author{Arne Brataas}
\affiliation{Department of Physics, Norwegian University of Science
  and Technology, N-7491 Trondheim, Norway}

\title{Spontaneous-Symmetry-Breaking Mechanism of Adiabatic Pumping}

\begin{abstract}
We consider heterostructures consisting of regions with a continuous symmetry in contact with regions wherein the symmetry is spontaneously broken. The low-frequency dynamics of the corresponding order parameter are shown to induce nonequilibrium transport, a ``pumping,'' out of the symmetry-broken regions, which is governed by the generator of the broken-symmetry operator. This pumping damps Goldstone-mode excitations and transfers them beyond traditional (static) proximity length scales. Our general conclusions are discussed for specific examples of order parameters in helimagnets, charge/spin-density waves, superconductors, and ferromagnets. We carry out a detailed calculation of such pumping for spiral magnetic orders in helimagnets possessing a duality in the representation of its symmetry-broken states.
\end{abstract}

\date{\today}
\maketitle


Transport in hybrid correlated-electron systems displays a variety of exciting physical phenomena that are governed by collective degrees of freedom. Examples include supercurrents in Josephson junctions and persistent spin currents in exchange-coupled ferromagnetic spin valves. Interesting effects can also occur when the nonequilibrium quasiparticle transport is initiated and manipulated by controlling the collective degrees of freedom. We show in the following that such nonequilibrium transport should be generated by any symmetry-breaking (SB) order parameter (OP) that is varied by external fields. As a specific example of pedagogical value, we focus on the spin and momentum pumping induced by sliding helimagnets in contact with normal metals.

Consider an ``island'' characterized by a local finite order parameter $\phi(\mathbf{r})$ describing some spontaneous SB embedded into a ``host'' with vanishing $\phi$. Electron correlations maintaining the finite OP in the island can penetrate into the host giving rise to equilibrium proximity effects in hybrid structures. We wish to study the nonequilibrium case when the time-dependent OP $\phi(\mathbf{r},t)$ is driven by an external field. It will be shown that the host quasiparticles respond to the order parameter dynamics close to the SB region, generating nonequilibrium transport that can propagate beyond the static proximity length scales. We develop a general phenomenological theory of dynamic coupling between the host and the island in terms of nonequilibrium pumping generated by the OP variation. This pumping generically exerts reaction forces that can be important for the OP dynamics. For demonstration purposes, we consider specific examples of helimagnets (HM), charge/spin-density waves (CDW/SDW), superconductors (SC), and ferromagnets (FM) in contact with normal metals (NM). Our conclusions may find applications in various areas of physics. 

Spin emission by precessing ferromagnets is already a well-studied phenomenon, which has recently attracted a renewed interest, leading to many useful insights after it was formulated as a parametric-pumping process \cite{Tserkovnyak:rmp}. In Ref.~\cite{Brataas:prl04}, we showed that spin and charge pumping by magnetic superconductors reflect their spin-pairing symmetry in experimentally-observable quantities. Here we show that the mean-field dynamics of \textit{any} symmetry-breaking OP generate nonequilibrium quasiparticle transport and propose an approach to simplify its computation. We apply the theory to helimagnets and charge/spin-density waves, as well as discuss the already studied cases of ferromagnets and superconductors in the context of a generalized formalism. To this end, let us consider an arbitrary SB island with partition function expressed as a functional integral over the OP field \cite{Negele}: $Z=\int\mathcal{D}[\phi(\mathbf{r})]\exp[-\beta\{\mathcal{F}[\phi(\mathbf{r})]+\int d\mathbf{r}\phi(\mathbf{r})\mathcal{H}(\mathbf{r})\}]$, where $\beta$ is the inverse temperature, $\mathcal{F}$ is an effective free-energy functional, and $\mathcal{H}(\mathbf{r})$ is an external field coupled to the OP.  Suppose the free energy $\mathcal{F}$ is invariant under a continuous transformation group $U(\theta)$ parametrized by $\theta$ [$U(0)=1$], $\mathcal{F}[U\phi]=\mathcal{F}[\phi]$, defining a symmetry that is broken by the states with nonvanishing OP $\phi$. In the absence of the external field $\mathcal{H}$, $U$ applied to a reference ground state $\phi_0$ generates a manifold $\Phi$ of states minimizing $\mathcal{F}[\phi]$: $\phi_\theta(\mathbf{r})=U(\theta)\phi_0(\mathbf{r})$. We are interested in the symmetry subgroup with nontrivial [apart from $U(0)$] and unique transformations on $\phi_0$: $U(\theta)\phi_0=U(\theta^\prime)\phi_0$ only and only if $\theta=\theta^\prime$. A slowly-varying $\mathcal{H}(\mathbf{r},t)$ can adiabatically move $\phi[\mathcal{H}(\mathbf{r})]$ through $\Phi$, giving us a handle to control the dynamics of the OP. Alternatively, a small oscillating component of $\mathcal{H}$ on top of a larger static component $\mathcal{H}_0$ can set off a resonant motion of $\phi$ near $\phi[\mathcal{H}_0(\mathbf{r})]$, as in the case of the ferromagnetic resonance.

The essential ingredient in our model is an effective mean-field Hamiltonian experienced by the host, which is modulated by the time-dependent OP $\phi(t)$ in the island. We view the SB island as a scatterer perturbing the state of the host. Coupling to the host also affects its properties, which should be taken into account self consistently. We wish to consider the host response to a time-dependent Hamiltonian $H[\phi(\mathbf{r},t)]$. To be specific, we consider an effective Hamiltonian $H=H_0+H^\prime$ describing electrons in a normal Fermi liquid, which are perturbed by the presence of an SB island. $H_0$ is an unperturbed Hamiltonian and $H^\prime[\phi(\mathbf{r})]$ is a coupling to the SB island. At sufficiently low temperatures, the island is characterized by a given position $\phi_\theta(\mathbf{r})$ on the manifold $\Phi$ of the ground states, which is controlled by the small external field $h$ and is labeled by $\theta$ defining a transformation from a certain reference state $\phi_0(\mathbf{r})$. The Hamiltonian is also labeled by $\theta$ correspondingly: $H^\prime(\theta)=H^\prime[\phi_\theta(\mathbf{r})]$, with each $\theta$ assumed to define a unique $H^\prime$. Suppose $H^\prime(\theta)$ can be generated from $H^\prime(0)$ by a certain unitary $\theta$-dependent transformation on the host degrees of freedom, and $H_0$ is invariant under this transformation,
\begin{equation}
H^\prime(\theta)=U(\theta)H^\prime(0)U^\dagger(\theta)\,,\,\,\left[H_0,U(\theta)\right]=0,
\label{uhu}
\end{equation}
where we naturally use the same symbol $U$.

\begin{figure}[pth]
\includegraphics[width=0.85\linewidth,clip=]{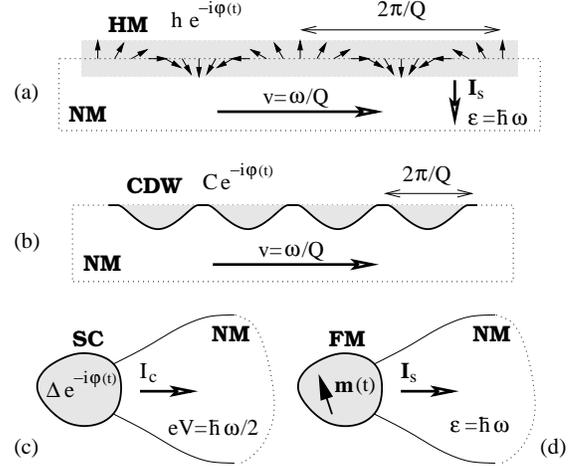}
\caption{Schematics of SB heterostructures. (a) HM dynamics can be viewed either as a rotating or sliding spiral order with frequency $\omega=\partial_t\varphi$ and wavevector $2\pi/Q$. These dynamics induce both momentum and spin flow into the normal metal. (b) A sliding CDW felt by an NM host that locks to its motion. CDW phase variation translates into velocity $v=\partial_t\varphi/Q$. The dynamics thus pump momentum. (c) Charge current $I_c$ pumped by the varying pairing-gap phase $\varphi(t)$ induces a charge imbalance in the normal metal corresponding to the voltage $V=(\hbar/2e)\partial_t\varphi$. (d) Spin pumping $\mathbf{I}_s$ by the rotating magnetization direction $\mathbf{m}(t)$ establishes spin splitting $\varepsilon=\hbar|\mathbf{m}\times\partial_t\mathbf{m}|$ along the rotation axis.}
\label{scheme}
\end{figure}

\begin{figure}[pth]
\includegraphics[width=0.9\linewidth,clip=]{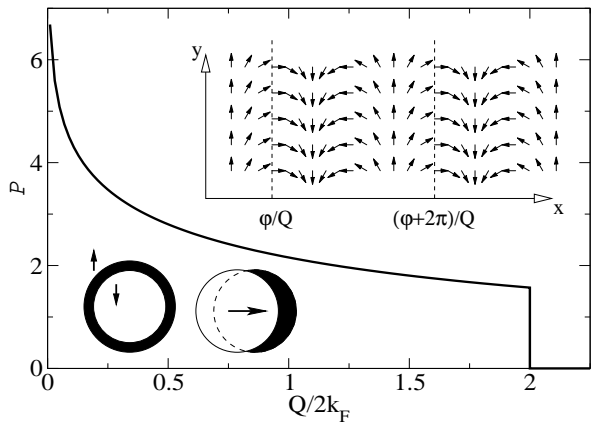}
\caption{Main panel: $\mathcal{P}$ defining the HM pumping, Eq.~(\ref{dSz}), as a function of $Q/2k_F$ (with the $y$ axis suppressed in the calculation). Upper inset: Spiral magnetic order in the helimagnet. Arrows show the local exchange field in the $xy$ plane for $h(\mathbf{r})=h_0\delta(z)e^{-i\varphi}$ and $\mathbf{Q}=Q\mathbf{\hat{x}}$, where $h_0$ is real-valued and $Q>0$. Phase $\varphi$ fixes the HM position in the $x$ direction. $\varphi$ can be varied by either spatial translation along the $x$ axis or spin rotation around the $z$ axis normal to the $xy$ plane. Lower inset: Two distinct stationary states of the normal metal corresponding to the HM dynamics. Left is the spin-polarized Fermi gas with energy splitting $\varepsilon=\hbar\omega$ along $z$ and right is the Fermi gas drifting along $x$ with velocity $v=\omega/Q$.}
\label{hm}
\end{figure}

To give more substance to our discussion, we consider several concrete examples shown in Fig.~\ref{scheme}. The Hamiltonian for host electrons scattering off the SB island and locally experiencing its OP is given in these examples by $H^\prime=\int d\mathbf{r}$ over the integrands
\begin{align}
&{\rm (a)}\,\,\Psi^\dagger_\uparrow\Psi_\downarrow(\mathbf{r})h(\mathbf{r})e^{i\mathbf{Q}\cdot\mathbf{r}}\,,\,{\rm (b)}\,\,\Psi^\dagger_\alpha\Psi_\alpha(\mathbf{r})C(\mathbf{r})e^{i\mathbf{Q}\cdot\mathbf{r}}\,,\nonumber\\
&{\rm (c)}\,\,\Psi_{\uparrow}^\dagger(\mathbf{r})\Psi_{\downarrow}^\dagger(\mathbf{r})\Delta(\mathbf{r})\,,\,{\rm (d)}\,\,\Psi_\alpha^\dagger\Psi_{\alpha^\prime}(\mathbf{r})\left[\boldsymbol{\sigma}_{\alpha\alpha^\prime}\cdot\boldsymbol{\epsilon}_{\text{xc}}(\mathbf{r})\right]
\label{H}
\end{align}
plus Hermitian conjugates in (a), (b), and (c), and the sum is implied over repeated indices. Here $\Psi_\alpha$'s are spin-$\alpha$ electron field operators, $h(\mathbf{r})$, $C(\mathbf{r})$, and $\Delta(\mathbf{r})$ are complex-valued HM, CDW, and SC gap functions, respectively, $\boldsymbol{\epsilon}_{\text{xc}}$ is the local ferromagnetic exchange field, and $\boldsymbol{\sigma}$ is a vector of the Pauli matrices. We take $C(\mathbf{r})$ to be uniform along the direction of $\mathbf{Q}$. The SDW Hamiltonian can be obtained from that of the CDW by adding an extra factor $(-1)^\alpha$ under the sum. The CDW can result from a Peierls instability in quasi-one-dimensional electron gases, leading to $Q=2k_F$ in terms of the Fermi wave vector $k_F$ \cite{Visscher:prb96}. The OP's, which self-consistently define $h$, $C$, $\Delta$, and $\boldsymbol{\epsilon}_{\text{xc}}$ experienced by quasiparticles, are given respectively by the HM spiral order $\langle\Psi_\downarrow^\dagger\Psi_\uparrow\rangle$ \cite{Heurich:prb03}, the CDW condensation amplitude $\langle\Psi_\alpha^\dagger\Psi_\alpha\rangle$ \cite{Visscher:prb96}, the Cooper-pair amplitude $\langle\Psi_{\uparrow}\Psi_{\downarrow}\rangle$, and the itinerant magnetization $\langle\Psi_\alpha^\dagger\boldsymbol{\sigma}_{\alpha\alpha^\prime}\Psi_{\alpha^\prime}\rangle$. We assume that the ferromagnetic exchange field can point in arbitrary direction $\mathbf{m}$, $\boldsymbol{\epsilon}_{\text{xc}}(\mathbf{r},t)=\epsilon_{\text{xc}}(\mathbf{r})\mathbf{m}$(t), breaking the spin-rotational invariance of the Hamiltonian $H$, and the gap functions can have arbitrary phase $\varphi$, $f(\mathbf{r},t)=f(\mathbf{r})e^{-i\varphi(t)}$, breaking the spin-rotational invariance around the $z$ axis for the helimagnet ($f\equiv h$), translational invariance for the charge/spin-density wave along $\mathbf{Q}$ ($f\equiv C$), and gauge invariance for the superconductor ($f\equiv\Delta$). Helimagnets whose gap function $h(\mathbf{r})$ is uniform along $\mathbf{Q}$ will be of a special interest for our discussion because of the dual representation of their symmetry-broken states: Any state determined by an overall phase of $h(\mathbf{r})$ can be obtained from a reference state with zero phase by either a rotation in spin space around the $z$ axis or a spatial translation along the direction of $\mathbf{Q}$, see the upper inset of Fig.~\ref{hm}.

The mean-field OP dynamics translate into the time dependence of $H^\prime$. Let us write the corresponding transformation along a given path parametrized by a single parameter $\theta(t)=\omega t$, where $\omega$ is the velocity along this path, as $U(t)=\exp(-iG\omega t)$, in terms of the generator $G=i\partial_\theta U(\theta)\vert_{\theta=0}$. Due to adiabaticity, any general motion can instantaneously be described by a constant-speed trajectory. The time dependence of $H=H_0+H^\prime$ is formally eliminated by applying the inverse transformation, $U^\dagger(t)$, to the host degrees of freedom and correspondingly adding a new term to the Hamiltonian, determined by the broken-symmetry generator:
\begin{equation}
H(t)\rightarrow H(0)-\hbar\omega G\,,
\label{Gen}
\end{equation}
subject to Eqs.~(\ref{uhu}). This can be loosely interpreted as follows: The effective Hamiltonian experienced by the host medium due to the dynamic SB island is time independent in the ``reference frame'' moving with the island. The additional term $\hbar\omega G$ in Eq.~(\ref{Gen}) reflects the inertial forces in the moving frame of reference and is given in our toy models, Eqs.~(\ref{H}), respectively by
\begin{align}
&{\rm (a)}\,\,S_z\partial_t\varphi\,\,{\rm or}\,\,(\mathbf{P}\cdot\mathbf{Q}/Q^2)\partial_t\varphi\,,\,{\rm (b)}\,\,(\mathbf{P}\cdot\mathbf{Q}/Q^2)\partial_t\varphi\,,\nonumber\\
&{\rm (c)}\,\,(\hbar N/2)\partial_t\varphi\,,\,{\rm (d)}\,\,\mathbf{S}\cdot(\mathbf{m}\times\partial_t\mathbf{m})\,,
\label{G}
\end{align}
where $\mathbf{S}$ is the total-spin operator, $\mathbf{P}$ is the total-momentum operator, and $N$ is the total-particle-number operator. The example of spiral order is special in that the time dependence of the Hamiltonian (\ref{H}a) can be removed by two different transformations: either rotation in spin space or translation in real space [if $\mathbf{Q}\cdot\partial_\mathbf{r}h(\mathbf{r})$=0]. The translation generator (\ref{G}b) can be removed by a gauge transformation (assuming the usual nonrelativistic Hamiltonian with kinetic energy $\mathbf{p}^2/2m$, per particle with momentum $\mathbf{p}$, and electron-electron interactions), which is consistent with the Galilean invariance dictating that there should be no inertial forces in a reference frame moving linearly at a constant speed.

If $U$ is a global-symmetry operator for the entire host medium, the new term in the Hamiltonian (\ref{Gen}) leads to a stationary state that could be different from that of $H_0$ even very far from the SB island. The simplest example is the CDW: The time dependence of the gap-function phase $\varphi(t)=\omega t$ corresponds to sliding of the effective potential experienced by the electrons at velocity $v=\omega/Q$ along the direction given by $\mathbf{Q}$, see Fig.~\ref{scheme}(b). Due to the gauge invariance, eigenstates of the transformed Hamiltonian $H(0)-vP_Q$, see Eq.~(\ref{G}b), can be obtained from the eigenstates of $H(0)$ by shifting the momenta of all electrons by $k=mv/\hbar$ along $\mathbf{Q}$. Transforming them back into the ``laboratory frame of reference'' shifts the wavefunctions in space without affecting the momenta, and we thus see that the entire host is sliding with velocity $v$ locked with the CDW, as can be naively expected. If the CDW starts slowly accelerating in the rest frame, the host will acquire momentum by ``pumping'' of finite-momentum electron-hole pairs from the CDW condensate that originates near the CDW region and propagates across the sample. An SDW can similarly pump momentum-carrying electron-hole pairs, an SC pumps Cooper pairs, an FM pumps spin-polarized electron-hole pairs, and an HM pumps both spin-polarized and momentum-carrying electron-hole pairs, as explained in the following.

Before proceeding, we point out that the spin-pumping physics \cite{Tserkovnyak:rmp} can be recast as merely another example of the SB mean-field pumping: According to Eq.~(\ref{G}d), magnetic direction  $\mathbf{m}(t)$ rotating with frequency $\omega$ spin polarizes the nonmagnetic structure in contact with the ferromagnet. In steady state, the nonmagnetic region is spin split along the rotation axis by energy $\hbar\omega$. This stationary spin accumulation, in turn, interacts with the FM island inducing spin currents from the NM region, which have to cancel the pumped spin currents \cite{Tserkovnyak:rmp}. Knowing a steady-state solution and calculating the dc transport corresponding to the nonequilibrium state of the system thus allows one to find the pumping circumventing the time-dependent transport problem. Another simple example that was already considered in literature \cite{Brataas:prl04} is the superconducting island in Ohmic contact with a normal metal, see Fig.~\ref{scheme}(d). The time-dependent gap function $\Delta(0)e^{-i\varphi(t)}$ effectively induces a voltage imbalance across the junction, determined by the Josephson equation.

It is instructive to discuss in some detail the HM$\mid$NM dynamic system. The upper inset of Fig.~\ref{hm} depicts the helimagnet with $h(\mathbf{r},t)=h_0\delta(z)e^{-i\omega t}$ and $\mathbf{Q}=Q\mathbf{\hat{x}}$. Since the Hamiltonian (\ref{H}a) is translationally invariant along the $y$ direction, we suppress this axis and consider a two-dimensional transport problem in the $xz$ plane (with the HM magnetization lying in the $xy$ plane). When such a helimagnet is treated as a perturbation embedded into a free-electron gas, the spiral dynamics have the following stationary solutions corresponding to Eqs.~(\ref{G}a): a Fermi sea spin polarized along the $z$ axis by energy splitting of $\hbar\omega$ or a Fermi sea drifting with velocity $\omega/Q$ along the $x$ axis, see lower inset of Fig.~\ref{hm} (or a linear superposition of the two: e.g., spin-polarized by $\hbar\omega/2$ and drifting with velocity $\omega/2Q$). Using Fermi's Golden Rule, it is straightforward to calculate the backflow of the corresponding spin accumulation or momentum imbalance to obtain the nonequilibrium pumping. We compute that the spin-imbalance scattering must be compensated by the spin and momentum pumping (at zero temperature, per HM unit length)
\begin{equation}
dS_z/dt=\omega S_p\mathcal{P}(q)\,\,{\rm and}\,\,dP_x/dt=\omega S_p\mathcal{P}(q)Q\,,
\label{dSz}
\end{equation}
where $\mathcal{P}(q)=\int_0^{\cos^{-1}(q-1)}d\zeta/\sqrt{1-(\cos\zeta-q)^2}$ for $0<q<2$ and $\mathcal{P}(q)=0$ for $q\geq2$, in terms of $q=|Q/k_F|$, and the integration is performed over the angle of incidence $\zeta$ with respect to the $x$ axis. $S_p=|h_0|^2m^2/(\pi^2\hbar^2k_F)$, where $m$ is the electron mass and $k_F$ is the Fermi wavevector. [It is straightforward to generalize Eqs.~(\ref{dSz}) to a layered geometry in three dimensions by integrating over the phase space along the $y$ axis.] We plot the pumping strength $\mathcal{P}$ dependence on $Q$ (which can be expressed in terms of elliptic integrals) in the main panel of Fig.~\ref{hm}. In the long-wavelength limit, $Q\to0$, when the HM becomes a uniform $\delta$-function FM, $\mathcal{P}$ diverges logarithmically. Mathematically, this is anticipated since $\int d\zeta/\sin\zeta=\ln[\tan(\zeta/2)]$ diverges at $\zeta\to0,\pi$. Physically, it means that spin pumping by a $\delta$-function ferromagnet is not a smooth function of the exchange field $h_0$ at $h_0=0$, so that the perturbation expansion breaks down. This problem has been already encountered in literature \cite{Simanek:prb03} and can be dealt with by properly including a cutoff for electrons that are incident on the magnetic layer at a shallow angle $\zeta\sim h_0k_F/2\varepsilon_F$. For short wavelengths, $Q\to2k_F$ (but $Q<2k_F$), pumping approaches a finite value $\mathcal{P}=\pi/2$, although the integration range shrinks to zero. At $Q>2k_F$, the Golden-Rule integration range vanishes and so does pumping, exhibiting a step at $Q=2k_F$. For a three-dimensional system, however, pumping vanishes at $Q\to2k_F$ and there is no step in $\mathcal{P}$, since only the largest ``slice'' of the Fermi gas in the $k_xk_z$ plane with $k_y=0$ gives a finite contribution.

For pumping corresponding to the drifting stationary solution, we obtain Eqs.~(\ref{dSz}) but with the integral $\mathcal{P}(q)=(2/q)\int_0^{\cos^{-1}(q-1)}d\zeta\cos\zeta/\sqrt{1-(\cos\zeta-q)^2}$ for $0<q<2$, which at first appears to be different from that in Eqs.~(\ref{dSz}). It turns out, however, that the integrals are numerically indistinguishable (although we were not able to prove it analytically). The pumped flows thus do not depend on the choice of a stationary solution, as expected to lowest order in frequency. It is however not clear which of the various stationary states is established in practice. Unlike pumping, this depends on the properties of the normal metal away from the HM$\mid$NM contact. If, for example, the electron gas cannot drift with the helimagnet because of boundaries or disorder scattering, HM dynamics will induce the spin accumulation $\hbar\omega$ along the $z$ axis, which must cancel the nonequilibrium spin pumping, in the absence of spin-relaxation processes. If, on the other hand, NM electron spins relax without momentum scattering, the electron gas will drift with velocity $\omega/Q$ along the $x$ axis. 

HM dynamics assumed in our theory can be excited by a magnetic field $h_z$ along the $z$ axis. In the case of an isolated dissipationless helimagnet, $h_z$ induces HM dynamics $\partial_t\varphi=\gamma h_z(t)$, where $\gamma$ is minus the gyromagnetic ratio. In order to see that these dynamics dissipate energy in the presence of an NM \textit{reservoir}, we calculate the imaginary part of the magnetic response function along the $z$ axis:
\begin{equation}
\mbox{Im}\chi(\omega)=\gamma^2S_p\omega/\left[(\gamma^2S_pK)^2+\omega^2\right]\,,
\end{equation}
where $K$ is the easy-plane anisotropy constant, assuming the HM prefers to lie in the $xy$ plane without an easy axis. Energy dissipation is an experimentally-observable manifestation of the pumping. Consequently, this demonstrates that the framework presented in this article can lead to easy calculation of experimentally-relevant quantities. Note that the spin pumping by a spiraling HM is different from the persistent spin currents predicted in Ref.~\cite{Heurich:prb03} for an equilibrated HM in the presence of a uniform magnetic field lying in the magnetization plane.

Let us now summarize a general approach for calculating the adiabatic pumping by any time-dependent Hamiltonian breaking a symmetry of the system: 1) The broken-symmetry operator $U$ satisfying Eqs.~(\ref{uhu}) is identified, 2) the system is transformed into the moving reference frame and solved for the stationary state after adding the new term in the Hamiltonian, which is proportional to the generator of $U$, Eq.~(\ref{Gen}), 3) the stationary nonequilibrium configuration induced by the additional term in the transformed Hamiltonian, and the corresponding backflow transport, are calculated in the laboratory frame. As this transport must be canceled by the pumping in the steady state, the procedure automatically gives (minus) the pumping due to the time-dependent Hamiltonian $H^\prime(t)$. This assumes that the total transport is given by the sum of the pumped current generated by the time-dependent OP and the backflow current induced by the nonequilibrium build up, which requires adiabaticity of the dynamics. We remark that it is not necessary that the steady state is reached in practice, since the pumping close to the SB island is not affected by the processes deep in the host. For a periodic variation of $H^\prime(t)$, the average pumping depends on the trajectory in the spontaneous-SB manifold. In particular, if the trajectory can be parametrized by a single periodic real-valued parameter, the net pumping over a cycle vanishes \cite{Brouwer:prb98}. In all of our examples, a constant-rate phase winding, $\varphi(t)=\omega t$, (or a steady magnetization rotation) can be described by two out-of-phase periodic parameters, $\cos(\varphi)$ and $\sin(\varphi)$, producing a finite net pumping.

Finally, we wish to comment on the reaction of the pumping on the OP dynamics. Since the process dissipates energy, the reaction of the medium on the SB island should damp the Goldstone modes restoring the broken symmetry. This damping can be calculated by first finding the pumped transport and then imposing the conservation laws in the classical equations of motion of the OP, as we did above for the helimagnets. A related phenomenon is the dynamic interaction between different SB regions: The nonequilibrium pumping induced by the OP dynamics can propagate from one SB island to another, damping Goldstone modes in one and exciting them in the other. Such dynamic interaction can be long ranged in comparison with the static interaction that requires quantum proximity. This formalism have already led to a considerable advancement in understanding of the ferromagnetic dynamics in hybrid structures \cite{Tserkovnyak:rmp}.

We acknowledge stimulating discussions with G. E. W. Bauer and B. I. Halperin. This work was supported in part by the Harvard Society of Fellows and by the Norwegian Research Council Nanomat program.


\begin{thebibliography}{6}
\expandafter\ifx\csname natexlab\endcsname\relax\def\natexlab#1{#1}\fi
\expandafter\ifx\csname bibnamefont\endcsname\relax
  \def\bibnamefont#1{#1}\fi
\expandafter\ifx\csname bibfnamefont\endcsname\relax
  \def\bibfnamefont#1{#1}\fi
\expandafter\ifx\csname citenamefont\endcsname\relax
  \def\citenamefont#1{#1}\fi
\expandafter\ifx\csname url\endcsname\relax
  \def\url#1{\texttt{#1}}\fi
\expandafter\ifx\csname urlprefix\endcsname\relax\def\urlprefix{URL }\fi
\providecommand{\bibinfo}[2]{#2}
\providecommand{\eprint}[2][]{\url{#2}}

\bibitem[{\citenamefont{Tserkovnyak et~al.}(2004)\citenamefont{Tserkovnyak,
  Brataas, Bauer, and Halperin}}]{Tserkovnyak:rmp}
For a review, see
  \bibinfo{author}{\bibfnamefont{Y.}~\bibnamefont{Tserkovnyak}},
  \bibinfo{author}{\bibfnamefont{A.}~\bibnamefont{Brataas}},
  \bibinfo{author}{\bibfnamefont{G.~E.~W.} \bibnamefont{Bauer}},
  \bibnamefont{and} \bibinfo{author}{\bibfnamefont{B.~I.}
  \bibnamefont{Halperin}},
  \bibinfo{note}{cond-mat/0409242 (unpublished)}.

\bibitem[{\citenamefont{Brataas and Tserkovnyak}(2004)}]{Brataas:prl04}
\bibinfo{author}{\bibfnamefont{A.}~\bibnamefont{Brataas}} \bibnamefont{and}
  \bibinfo{author}{\bibfnamefont{Y.}~\bibnamefont{Tserkovnyak}},
  \bibinfo{journal}{Phys. Rev. Lett.} \textbf{\bibinfo{volume}{93}},
  \bibinfo{pages}{087201} (\bibinfo{year}{2004}).

\bibitem[{\citenamefont{Negele and Orland}(1988)}]{Negele}
\bibinfo{author}{\bibfnamefont{J.~W.} \bibnamefont{Negele}} \bibnamefont{and}
  \bibinfo{author}{\bibfnamefont{H.}~\bibnamefont{Orland}},
  \emph{\bibinfo{title}{Quantum Many-Particle Systems}}
  (\bibinfo{publisher}{Perseus Books}, \bibinfo{address}{Reading,
  Massachusetts}, \bibinfo{year}{1988}).

\bibitem[{\citenamefont{Visscher and Bauer}(1996)}]{Visscher:prb96}
\bibinfo{author}{\bibfnamefont{M.~I.} \bibnamefont{Visscher}} \bibnamefont{and}
  \bibinfo{author}{\bibfnamefont{G.~E.~W.} \bibnamefont{Bauer}},
  \bibinfo{journal}{Phys. Rev. B} \textbf{\bibinfo{volume}{54}},
  \bibinfo{pages}{2798} (\bibinfo{year}{1996}).

\bibitem[{\citenamefont{Heurich et~al.}(2003)\citenamefont{Heurich,
  K{\"{o}}nig, and MacDonald}}]{Heurich:prb03}
\bibinfo{author}{\bibfnamefont{J.}~\bibnamefont{Heurich}},
  \bibinfo{author}{\bibfnamefont{J.}~\bibnamefont{K{\"{o}}nig}},
  \bibnamefont{and} \bibinfo{author}{\bibfnamefont{A.~H.}
  \bibnamefont{MacDonald}}, \bibinfo{journal}{Phys. Rev. B}
  \textbf{\bibinfo{volume}{68}}, \bibinfo{pages}{064406}
  (\bibinfo{year}{2003}).

\bibitem[{\citenamefont{Simanek}(2003)}]{Simanek:prb03}
\bibinfo{author}{\bibfnamefont{E.}~\bibnamefont{\v{S}im\'anek}} and
  \bibinfo{author}{\bibfnamefont{B.}~\bibnamefont{Heinrich}},
  \bibinfo{journal}{Phys. Rev. B} \textbf{\bibinfo{volume}{67}},
  \bibinfo{pages}{144418} (\bibinfo{year}{2003}).

\bibitem[{\citenamefont{Brouwer}(1998)}]{Brouwer:prb98}
\bibinfo{author}{\bibfnamefont{P.~W.} \bibnamefont{Brouwer}},
  \bibinfo{journal}{Phys. Rev. B} \textbf{\bibinfo{volume}{58}},
  \bibinfo{pages}{R10135} (\bibinfo{year}{1998}).

\end{thebibliography}
\end{document}